**Soliton Transitions Mediated by Skin-Mode Localization and Band Nonreciprocity**

Shanyue Li[1], Mengying Hu[1], Jing Lin[1], Chen Fang[2,3], Zhensheng Tao[1], and Kun Ding[1]†

[1]*Department of Physics, State Key Laboratory of Surface Physics, and Key Laboratory of Micro and Nano Photonic Structures (Ministry of Education), Fudan University, Shanghai 200438, China*

[2]*Beijing National Laboratory for Condensed Matter Physics, and Institute of Physics, Chinese Academy of Sciences, Beijing 100190, China*

[3]*Kavli Institute for Theoretical Sciences, Chinese Academy of Sciences, Beijing 100190, China*

† Corresponding E-mail: kunding@fudan.edu.cn

**Abstract**

Solitons, typically resulting from a competition between band dispersion and nonlinearity, occur in lattices featuring the non-Hermitian skin effect as nonlinearity increases, accompanied by a transition in localization from linear skin modes to solitons. However, localization does not disentangle the role of skin modes in the soliton formation from that of band dispersion. Here, in such lattices, we uncover two distinct soliton phases, skin-mode-assisted solitons (SMASs) and nonreciprocity-dressed solitons (NRDSs). Rooted in fundamentally different mechanisms, SMASs originate from skin effect, while NRDSs stem from band nonreciprocity, each exhibiting unique spatial profiles. Using a stacked Su-Schrieffer-Heeger-like model as a prototype, we delineate the phase diagram of SMASs and NRDSs, each having clear phase boundaries. To interpret them, we formulate a Wannier-function-based nonlinear Hamiltonian, showing that soliton formation depends critically on how skin-mode localization and band nonreciprocity suppress or enhance wave dispersion. For SMASs, skin-mode localization reduces wave broadening at the localization sites, thereby lowering the formation threshold. This soliton phase is observable from edge dynamics and accompanied by a dynamical stability reentrance when transitioning from linear skin modes. In contrast, NRDSs, as well as their thresholds, originate from bulk band nonreciprocity and persist under periodic boundary conditions. Our framework offers predictive tools for characterizing and engineering solitons in experimentally realizable non-Hermitian systems, spanning optics to mechanics.

***Introduction***—Nonlinear wave phenomena are fundamental to diverse physical systems, with solitons as robust and self-localized wave packets, playing a pivotal role in applications such as information transport, nonlinear optics, and signal processing [1–4]. Remarkable advances have been made in harnessing solitons within discrete lattice systems, uncovering myriad phenomena where soliton formation and propagation stem from the interplay between nonlinear self-focusing, band dispersion, and topology [5–22]. The advent of non-Hermitian lattices has revolutionized band dispersion by introducing complex energy, giving rise to band nonreciprocity under periodic boundary conditions (PBCs) [23–29]. The ensuing topologically nontrivial point gap results in skin-mode localization, known as the non-Hermitian skin effect (NHSE), which indicates that eigenstates accumulate exponentially at system boundaries under open boundary conditions (OBCs) in the linear regime [30–44]. When interacting with nonlinearity, the NHSE inherently introduces unprecedented factors into the soliton landscape, offering possibilities that are fundamentally inaccessible in Hermitian settings [45–57].

Recent studies show that, under single-channel excitation, increasing nonlinearity can transform skin modes into trapped solitons, with the excitation site strongly influencing soliton localization [45–47]. While such transitions have been experimentally observed [48], determining the critical nonlinearity at which solitons emerge as eigenstates remains challenging from a dynamical perspective alone, as linear skin modes tend to dominate long-time dynamics [45–47]. Existing theoretical treatments [46–49] and the terminology they adopt, such as nonlinear skin modes [49] and edge/bulk skin solitons [46–48], emphasize that such a localization transition stems from the mutual effect of skin-mode localization and band nonreciprocity [45–49]. Although they often coexist, these are physically distinct ingredients, especially when lattice intricacy increases [58–61]. Thus, it is natural to inquire into the existence of soliton phases that are determined by their respective mechanisms, and subsequently, the feasibility of predicting the transition threshold.

Building on this foundation, we employ a stacked Su-Schrieffer-Heeger-like model as a prototype system that incorporates both nonreciprocal hopping and third-order Kerr nonlinearity (top of Fig. 1). By mapping out the phase diagram, we reveal two distinct phases of solitons, with their phase boundaries clearly defined: skin-mode-assisted solitons (SMASs) and nonreciprocity-dressed solitons (NRDSs). As their names suggest, SMASs (green profile in Fig. 1) are tightly associated with skin-mode localization, while NRDSs (red profile in Fig. 1) originate from band nonreciprocity, both leaving distinct signatures in their spatial profiles. To quantitatively determine their phase boundaries and formation mechanisms, we construct a Wannier-function-based nonlinear Hamiltonian. By retaining dominant Wannier functions to work out the analytical solutions, we demonstrate that the nonlinearity threshold $g_s$ ($g_n$) for SMASs (NRDSs) is governed by the degree to which skin-mode localization (band nonreciprocity) suppresses or enhances wave dispersion. Both soliton phases are dynamically accessible, although the transition from skin modes to SMASs is concomitant with dynamical instability. By clarifying the dominant mechanisms of each soliton phase, our framework enables predictive control over soliton formation and transitions in non-Hermitian lattices.

*Skin-mode-assisted solitons and nonreciprocity-dressed solitons*—To inspect the occurrence of SMASs and NRDSs, we consider a finite lattice under OBCs (Fig. 1), and its unit cell is shown in Fig. 2(a). The system features nonreciprocal inter-cell hopping as $\kappa_R = \kappa(1-\gamma)$ and $\kappa_L = \kappa(1+\gamma)$, where $\gamma \in [0,1]$. The nonlinearity is of the Kerr type, characterized by a nonlinearity strength parameter $g$, with $g < 0$ corresponding to the focusing regime. The real-space eigenvalue equation is then formally expressed as

$$E\psi_n = H_{nm}\psi_m + g|\psi_n|^2\psi_n, \tag{1}$$

where $\psi_n$ is the wavefunction at site $\mathbf{n} = (n_x, n_y)$ with $n_x = 1,2,\cdots,N_x$ and $n_y = 1,2,\cdots,N_y$. $H_{nm}$ is the linear Hamiltonian, and $E$ denotes the nonlinear eigenenergy. The number of unit cells along the $x$ and $y$ directions is $N_x/2$ and $N_y$, respectively, with odd (even) $n_x$ corresponding to sublattice A (B).

Before proceeding with full solutions of Eq. (1), we first examine two limiting cases: the linear limit and the molecular limit. Setting $g = 0$ recovers the linear regime, where NHSE necessarily arises when $\gamma \neq 0$. The shaded blue region in Fig. 2(c) indicates the OBC spectral range for $\gamma = 0.5$, and the skin mode with the lowest OBC energy, as shown by the dashed line in Fig. 2(b), is localized at the left boundary (see Sec. I in Ref. [62]). The molecular limit corresponds to the case with no inter-cell hoppings, i.e., $\kappa_R = \kappa_L = t_y = 0$ [63]. In this regime, Eq. (1) decouples into a series of identical nonlinear equation systems, each consisting of two coupled equations. Solving these decoupled systems yields the spectra depicted by the solid black and grey lines in Fig. 2(c), with corresponding eigenstate profiles shown in the inset (see Sec. II in Ref. [62]). Owing to the underlying sublattices, two distinct types of nonlinear eigenstates emerge: type-I states, which are localized on a single site (gray line and inset), and type-II states, which are symmetrically localized across two sites (black line and inset).

With the above preparation, we now numerically solve Eq. (1) using an iterative method (see Sec. III in Ref. [62]). Since we are interested in solitons across the entire lattice, we use the solutions in the molecular limit as initial trials and record steady-state solutions [Im($E$) =

0] at all lattice sites. The resulting spectra for representative nonlinear eigenstates are displayed by various markers in Fig. 2(c).

In the weakly nonlinear region ($g \to 0$), blue circles in Fig. 2(c) indicate the spectra that start from the skin mode with the lowest OBC energy, with a typical $|\psi_n|$ shown in Fig. 2(b). Evidently, $|\psi_n|$ remains localized near the left boundary, closely resembling the linear skin mode. To verify this, we apply regular perturbation theory [49], treating the nonlinearity as a perturbation to the linear skin mode (see Sec. IV in Ref. [62]), and the obtained spectrum is shown by the solid green line in Fig. 2(c). A good agreement suggests that the nonlinear eigenstates, represented by blue circles, are essentially the same as linear skin modes, and thus, we refer to them as nonlinear perturbative skin modes (NPSMs).

As $g$ increases, the intersection of the green and black lines in Fig. 2(c) hints that nonlinearity is no longer a perturbation and solitons are beginning to emerge. Numerically, solitons are first found near the left boundary, where linear skin modes localize. Their spectra, shown as open triangles in Fig. 2(c), nearly follow the molecular limit. However, as displayed in Fig. 2(d), while these solitons remain confined to a single unit cell, their tails decay exponentially into the bulk and approach a finite value at the boundary, suggesting a strong connection to skin-mode localization. To highlight this connection, the bottom panel of Fig. 2(d) shows both the linear skin mode (dashed line) and the amplitude level $1/\sqrt{N}$ of a fully extended mode (shaded region), where $N$ is the total number of lattice sites. These solitons emerge from the NHSE-induced eigenstate localization and are thus referred to as SMASs. Figures 2(c) and 2(d) present two typical SMASs, where $SMAS_1$ (green triangles), localized at the leftmost unit cell, forms at a smaller $g$ than $SMAS_2$ (cyan triangles), which lies further into the bulk. The subscript used in SMAS denotes the unit-cell position where SMASs localize with respect to the left boundary. Hence, we define the nonlinearity threshold for the formation of SMAS at the localization site of the linear skin mode as $g_s$ [green dashed line in Fig. 2(c)].

For lattice sites away from NHSE localization regions, solitons also form. Their spectra are shown as red crosses in Fig. 2(c), with corresponding profiles displayed in Fig. 2(e). As expected, these solitons appear deep in the bulk and require a higher nonlinearity threshold than SMASs. Unlike SMASs, they decay exponentially on both sides, rendering the boundary irrelevant. Their profiles are highly asymmetric, with the decay length on the left being longer than that on the right, reflecting the influence of nonreciprocal hoppings rather than eigenstate localization of NHSE. We designate these as NRDSs, with their formation threshold denoted as $g_n$ [red dashed line in Fig. 2(c)]. Inspecting the NRDS profiles within a localization unit cell, we find that they retain the same types as those in the molecular limit [see insets of Fig. 2(c)], with their types labeled by the subscripts of NRDSs [top two panels in Fig. 2(e)]. As shown in Fig. 2(c), with incrementally increasing $|g|$, NRDS$_{II}$ emerges when $|g| > |g_n|$ and transitions to NRDS$_I$ at a critical value $g_t$ [amaranth dashed line in Fig. 2(c)]. While $g_s$ and $g_n$ capture the formation threshold of SMASs and NRDSs, respectively, $g_t$ is more tied to the underlying sublattices, as indicated by the molecular limit. We have used the subscript of $g$ to label the threshold for each soliton phase. Although Fig. 2 only shows the scenario for one particular $\gamma$, the intrinsic characteristics of $g_s$, $g_n$, and $g_t$ have been revealed, and hence, their dependence on $\gamma$ is expected to be distinct.

*Phase diagram*—To validate the soliton phases identified above, we repeat the analysis across various $\gamma$ and display the phase diagram in Fig. 3. NPSMs, SMASs, and NRDSs occupy distinct parameter regions, and NRDSs (SMASs) recover bulk (edge) solitons when $\gamma = 0$. The thresholds, $g_s$, $g_n$, and $g_t$ (solid lines), exhibit different dependencies on $\gamma$. Among them, $g_t$ shows minimal variation, indicating that this transition is almost independent of nonreciprocal hoppings. This aligns with the fact that NRDS types mirror those in the molecular limit, which predicts the transition at $g = 2t_x$ (see Sec. II in Ref. [62]). Despite slight deviations between $g_t$ and $-4$ ($= 2t_x$) resulting from residual inter-cell couplings, the overall agreement confirms the transition behavior.

The above discussion further suggests that eigenstate localization from the NHSE is irrelevant for NRDSs. This raises the question: is $g_n$, the formation threshold for NRDSs, also independent of skin-mode localization? We thereby replace OBCs with PBCs to eliminate such localization. Under PBCs, both NPSMs and SMASs, together with $g_s$, disappear and become delocalized modes, while NRDSs persist (see Sec. V in Ref. [62]). Notably, $g_n$ and $g_t$, marked by open squares and circles in Fig. 3, closely match their counterparts under OBCs. This agreement not only supports the connection between $g_t$ and molecular limit, but also indicates that $g_s$ is tightly linked to skin-mode localization, whereas $g_n$ is not.

To disclose the dominant factors in $g_s$ and $g_n$, we expand solutions to Eq. (1) in a Wannier basis (see Sec. VI in Ref. [62]) [21,22,64]

$$\psi_n = \sum_{\boldsymbol{R},\alpha} c_{\boldsymbol{R},\alpha} w^{\text{R}}_{\boldsymbol{R},\alpha,n}, \tag{2}$$

where $c_{\boldsymbol{R},\alpha}$ are expansion coefficients, $\boldsymbol{R} = (R_x, R_y)$ denotes the real-space lattice vector, $\alpha$ is the band index, and $w^{\text{R}}_{\boldsymbol{R},\alpha,n}$ are the Wannier functions constructed from PBC right eigenvectors (with superscript R; L for left). By substituting Eq. (2) into Eq. (1) and assuming nonlinear interactions are dominated by Wannier functions of the valence band at the same unit cell, we obtain the Wannier-function-based nonlinear Hamiltonian $H_{R,R'} c_{R'} = E c_{R'}$,

$$H_{R,R'} = \epsilon_{R,R'} + g W_{R,R,R,R}^{\alpha,\alpha,\alpha,\alpha} |c_R|^2 \delta_{R,R'}, \tag{3}$$

where $W_{R,R,R,R}^{\alpha,\alpha,\alpha,\alpha}$ is the four-function overlap integral

$$W_{R,R,R,R}^{\alpha,\alpha,\alpha,\alpha} = \sum_n (w_{R,\alpha,n}^{\mathrm{R}})^* w_{R,\alpha,n}^{\mathrm{L}} w_{R,\alpha,n}^{\mathrm{R}} w_{R,\alpha,n}^{\mathrm{R}}, \tag{4}$$

with $\alpha$ taken as the valence band. $\epsilon_{R,R'}$ reflects the real-space band dispersion, and, within the PBC setting, can be obtained by Fourier transforming the Bloch band dispersion $E_k$, i.e., $\epsilon_{R,R'} = \epsilon_{R'-R} = 2[\sum_k e^{-ik \cdot (R'-R)} E_k]/N$. OBCs can then subsequently be applied by setting the boundary elements in $\epsilon_{R'-R}$ to zero. To verify, we utilize Eq. (3) to examine SMAS$_1$ in Fig. 2(d) and NRDS$_{\mathrm{II}}$ in Fig. 2(e), as shown by open circles in Figs. 4(a) and 4(b). Excellent agreement with solid lines validates the Wannier-function-based nonlinear Hamiltonian. Accordingly, $g_s$ and $g_n$ in the phase diagram are successfully reproduced, indicated by open diamonds and pentagrams in Fig. 3. This accurate reproduction suggests that the formalism provides a powerful tool to uncover the underlying mechanisms.

Figures 4(c) and 4(d) depict expansion coefficients $c_R$ throughout the entire lattice corresponding to Figs. 4(a) and 4(b), which show that both SMASs and NRDSs can be reconstructed by using several Wannier functions. Therefore, we retain the two leading Wannier functions and work out $g_s$ and $g_n$ analytically (see Sec. VI in Ref. [62] for full analytical expressions). For $g_s$, we consider Wannier functions at the leftmost two-unit cells and find that $g_s \sim \epsilon_\rightarrow$, where $\epsilon_\rightarrow = \epsilon_{(-3,0),(-4,0)}$ denotes the rightward hopping, which is opposite to the skin-mode localization direction. In contrast, $g_n \sim \epsilon_\leftarrow$, where $\epsilon_\leftarrow$ is the leftward hopping aligned with the skin-mode localization direction. These formulas unveil distinct formation mechanisms for SMASs and NRDSs. Conventionally, solitons form when wave dispersion or broadening is balanced by nonlinear self-focusing. For NRDSs, wave dispersion becomes asymmetric and boosted along the dominant hopping direction [inset in Fig. 4(d)], making $g_n$ increase against $\gamma$. However, wave dispersion is suppressed at sites where skin modes localize due to NHSE, making it proportional to the delocalizing hopping amplitude of skin modes [inset in Fig. 4(c)], and hence, $g_s$ decreases as $\gamma$ ramps up. This intuitive understanding not only validates the nonlinear Hamiltonian, which clarifies distinct roles played by NHSE and nonreciprocity in the soliton formation, but also substantiates the phase diagram encompassing both soliton phases and their transition mechanisms.

***Stability reentrance and edge dynamics***—As NRDSs spatially differ from SMASs, probing NRDSs, along with the $g_n$ and $g_t$ transition, by bulk dynamics is straightforward (see Sec. VII in Ref. [62]), but distinguishing SMASs from NPSMs and linear skin modes has inevitable difficulties because they often appear at the same spatial locations. Hence, exciting the lattice deep in the bulk with an instantaneous source, a common protocol in bulk dynamics, is unfavorable due to the low excitation efficiency of SMASs and NPSMs [45]. The intrinsic difference in their localization lengths offers the possibility of identification, although the coexistence of two localization scales potentially induces dynamical instability [65].

Hence, we first perform a stability analysis by introducing perturbations to the obtained eigenstates and constructing the stability matrix $Z$ (see Appendix A for details). Evidently, only the transition from NPSMs to SMASs is accompanied by a stability gap, whereas other transitions are dynamically stable. Since both NPSMs and SMASs are accessible from the boundaries, we use edge dynamics to probe the $g_s$ transition by tracking the spatiotemporal evolution (see Appendix B for details). Such an observation not only offers a means to probe the phase transition but also necessitates distinguishing between NPSMs and SMASs, as this distinction fundamentally separates the nonlinear regime from the linear one.

***Discussions and conclusions***—All soliton phase transitions have been theoretically clarified by isolating the roles of skin-mode localization and band nonreciprocity. This achievement solidifies the introduced terminology, which reflects the underlying mechanisms purposefully and aptly. Since skin modes can occur at corners instead of edges in certain models, SMASs (NRDSs) emerge at the lattice sites where skin modes are localized (or not), highlighting the generality of our framework.

In summary, by deliberately configuring boundaries, we distinguish two soliton phases from numerically obtained nonlinear eigenstates supported in non-Hermitian lattices, dubbed SMAS and NRDS, which respectively owe to skin-mode localization and band nonreciprocity. Based on this classification, we acquire the phase diagram with clearly identified phase boundaries. To interpret and predict these transitions, we develop a Wannier-function-based nonlinear Hamiltonian that captures phase transitions and reveals the dominant mechanisms for each soliton phase. The formation of SMASs (NRDSs) roots in the suppression or enhancement of wave dispersion from skin-mode localization (band nonreciprocity). Additionally, we uncover a stability reentrance at the NPSM-SMAS transition and its dynamical accessibility from edge excitations. Given the experimental realization of non-Hermitian lattices with nonlinearity [48,66], our framework provides a practical tool for analyzing the interplay between skin modes and solitons, and potentially extends to gap solitons and nonlinear topological modes [5–22,50–55]. Finally, the anomalous dynamical behaviors of solitons in non-Hermitian lattices [45–47,67,68], combined with our stability analysis, suggest promising avenues for controlling solitons. Our results thus provide both conceptual insight and a strategy for soliton engineering in non-Hermitian systems.

**Acknowledgment**

We thank Prof. C. T. Chan, Prof. Yun Jing, Dr. Jia-Xin Zhong, and Dr. Kai Zhang for helpful discussions. This work is supported by the National Natural Science Foundation of China (12174072, 2021hwyq05, 12325404, 12188101), the National Key R&D Program of China (2022YFA1404701, 2022YFA1404500, 2023YFA1406704), and the Shanghai Science and Technology Innovation Action Plan (No. 24Z510205936).

*Appendix A: Stability analysis*—To perform the linear stability analysis [65], we write down the time-dependent equation governed by Eq. (1)

$$i\frac{d\Psi_n}{dt} = H_{nm}\Psi_m + g|\Psi_n|^2\Psi_n, \tag{A1}$$

where $\Psi_n(t)$ is in the time domain. Equation (A1) recovers Eq. (1) when $\Psi_n(t) \to e^{-iEt}\psi_n$. We then suppose perturbations to a given eigenstate $\psi_n$ as $\Psi_n(t) = [\psi_n + \epsilon v_n(t)]e^{-iEt}$, where $\epsilon \ll 1$. By substituting it into Eq. (A1) and assuming $v_n(t) = p_n e^{-i\lambda t} + q_n^* e^{i\lambda^* t}$, we obtain the stability matrix $Z$ with $\lambda$ and $(p_n \quad q_n)^{\mathrm{T}}$ being its eigenvalue and eigenvector. When all eigenvalues $\lambda$ of $Z$ are real, the corresponding mode is linearly stable; otherwise, it is unstable. We thus use the largest imaginary part of $\lambda$, denoted by $\mathrm{Im}[\lambda]_{\max}$, as a stability indicator. The magenta line in Fig. 2(c) and the color map in Fig. 5(a) show $\mathrm{Im}[\lambda]_{\max}$ for the NPSMs, SMAS, and NRDSs presented therein. Evidently, instability only appears near the transition from NPSMs to SMASs, indicating the presence of a stability gap near $g_s$.

To verify the results of stability analysis, we investigate system dynamics by using the eigenstate with perturbations as the initial excitation, i.e., $\Psi_n(t = 0) = \psi_n + \epsilon v_n(t = 0)$ with $v_n$ being the eigenvector corresponding to $\mathrm{Im}[\lambda]_{\max}$. Figures 5(b) to 5(d) display the evolution of one linear skin mode and two NPSMs, while Figs. 5(e) to 5(g) depict those for one SMAS and two NRDSs. Clearly, the transition from NPSMs to SMASs ($g_s$) undergoes a stability reentrance [Figs. 5(c) to 5(e)], while two soliton transitions, $g_n$ [Figs. 5(e) to 5(f)] and $g_t$ [Figs. 5(f) to 5(g)], do not experience an alteration in stability. This suggests that dynamical stability offers another means to probe soliton transitions in the phase diagram.

*Appendix B: Edge dynamics*—To demonstrate that both NPSMs and SMASs are accessible from edge dynamics, we excite the lattice at the boundary favored by both phases and record the evolution of wavefunctions, as shown in Figs. 6(a) and 6(b). We see that when $g$ is below $g_s$, no solitons are observed, while soliton features clearly appear when $|g| > |g_s|$. To quantify, we compute the time-averaged wavefunction amplitude

$$S_{\bm{n}} = \frac{1}{T/2} \int_{T/2}^{T} |\Psi_{\bm{n}}(t)| \mathrm{d}t, \tag{B1}$$

where $T$ is the total recorded evolution time [45]. The upper and lower panels of Figs. 6(c) and 6(d) display the distribution of $S_{\bm{n}}$ across the entire lattice and along the slice $n_y = 3$, respectively. Compared to the nonlinear eigenstates (blue and green lines), both the NPSM and SMAS$_1$ are clearly resolved, thereby confirming their accessibility via edge dynamics.

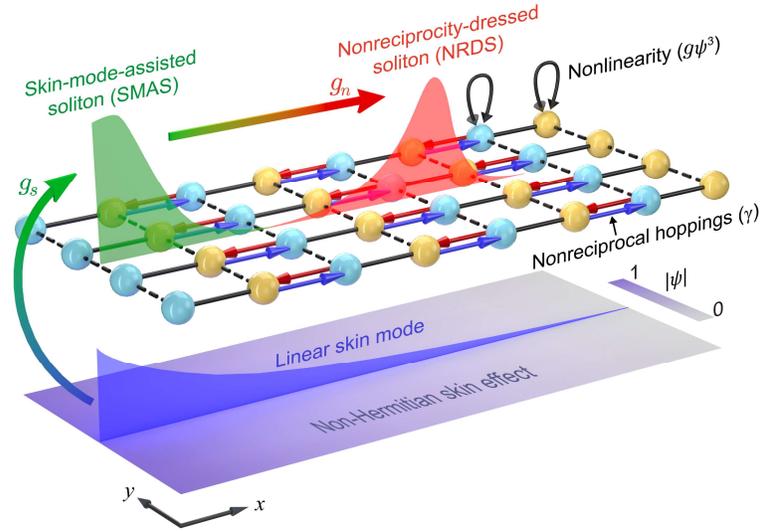

FIG. 1. Schematic illustration of transitions among skin modes and various solitons. The upper panel shows a two-dimensional lattice featuring nonreciprocal hoppings and nonlinearity, with their respective strengths characterized by $\gamma$ and $g$. Two arrows with gradient color sketch the sequential transitions from skin modes (blue) to SMASs (green) and subsequently to NRDSs (red), driven by variations in $g$.

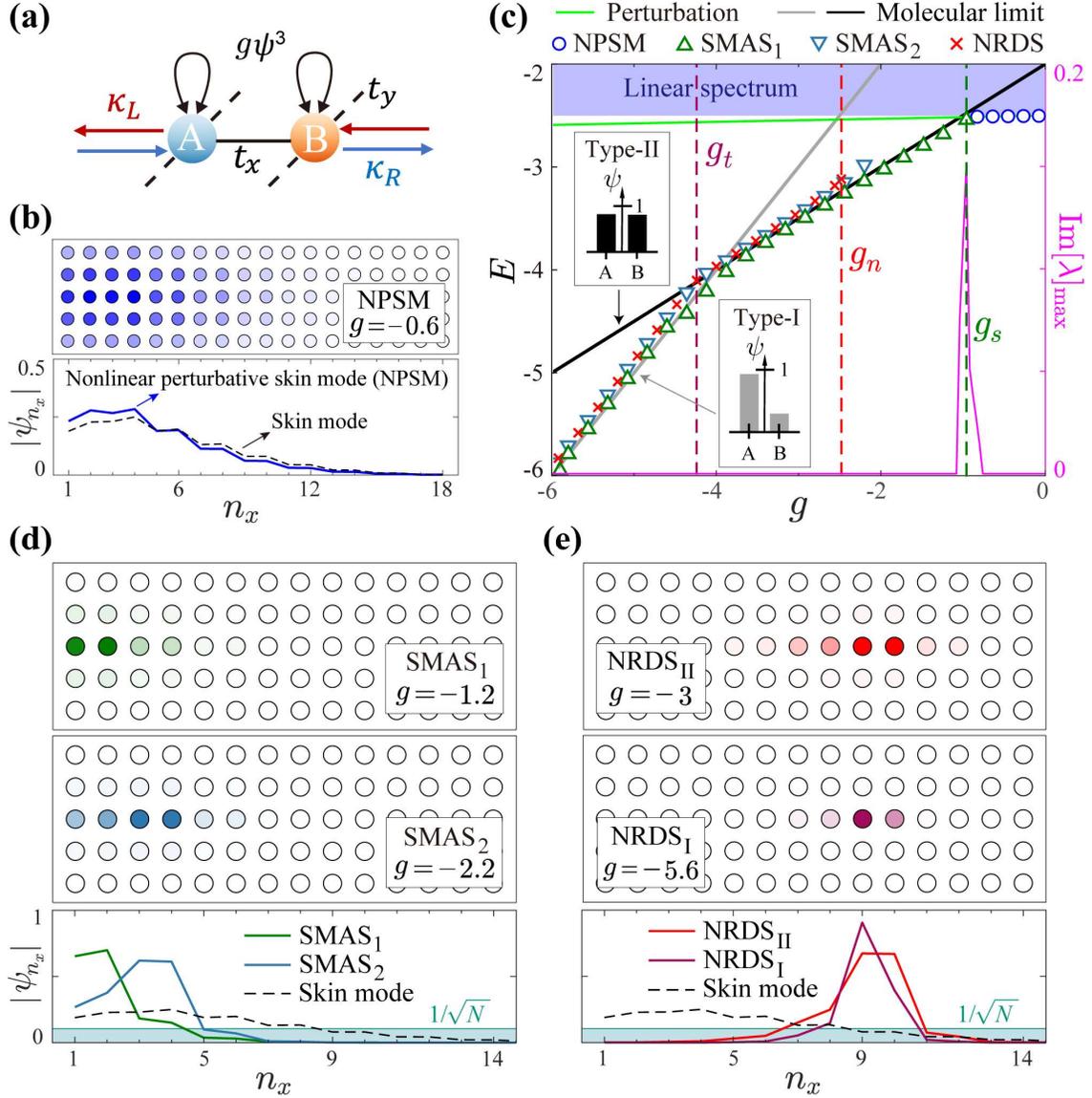

FIG. 2. (a) Unit cell with on-site Kerr nonlinearity. The intra-cell and $y$-direction inter-cell hoppings are reciprocal, denoted by $t_x$ and $t_y$, respectively. The $x$-direction inter-cell hopping is nonreciprocal, denoted by $\kappa_R$ (rightward) and $\kappa_L$ (leftward). (b) $|\psi_n|$ for an NPSM at $g = -0.6$ (upper panel). The lower panel shows $|\psi_{(n_x, n_y=3)}|$ of this NPSM (solid line) and a linear skin mode (dashed line). (c) Spectra of NPSMs (blue circles), SMASs (green and cyan triangles), and NRDSs (red cross markers). The dashed lines mark the transition thresholds, $g_s = -0.96$ (green), $g_n = -2.48$ (red), and $g_t = -4.24$ (amaranth). The black and gray lines indicate analytical spectra in the molecular limit, with insets showing the corresponding eigenstate profiles. The solid green line is obtained from perturbation theory. The magenta line exhibits $\text{Im}[\lambda]_{\max}$, the largest imaginary part of the eigenvalues of the stability matrix for the considered modes. (d) $|\psi_n|$ for SMAS$_1$ at $g = -1.2$ and SMAS$_2$ at $g = -2.2$ (upper panel). (e)

$|\psi_n|$ for NRDS$_{II}$ at $g = -3$ and NRDS$_I$ at $g = -5.6$ (upper panel). The solid lines in the lower panels of (d-e) show the corresponding $|\psi_{(n_x,n_y=3)}|$ for SMASs and NRDSs in the upper panels, while the dashed black lines represent linear skin modes with the shaded regions indicating fully extended modes. Panels (d,e) display one portion of the entire lattice. The color shading in the upper panels of (b,d,e) represents the amplitude $|\psi_n|$. All results are obtained under OBCs, with parameters $\kappa = -0.5$, $\gamma = 0.5$, $t_x = -2$, $t_y = -0.05$, $N_x = 18$, and $N_y = 5$.

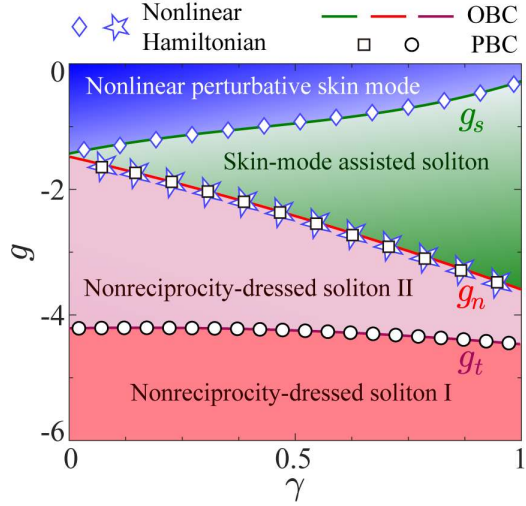

FIG. 3. Phase diagram in the $\gamma - g$ plane under OBCs. Each phase is labeled accordingly, with the transition boundaries $g_s$, $g_n$, and $g_t$ indicated by solid lines. Open squares and circles represent the $g_n$ and $g_t$ boundaries obtained under PBCs, respectively, while open diamonds and pentagrams indicate the values of $g_s$ and $g_n$ determined using the Wannier-function-based nonlinear Hamiltonian. The parameters used are identical to those in Fig. 2.

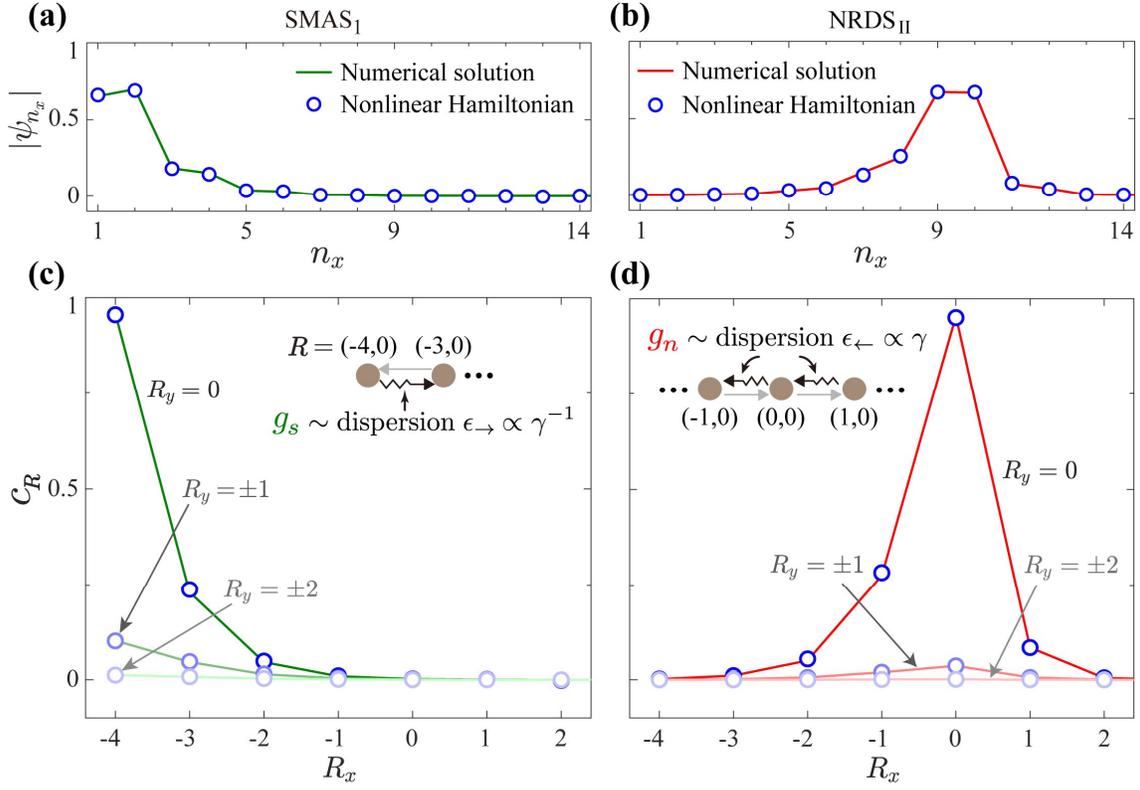

FIG. 4. (a)-(d) Comparison of $|\psi_{(n_x,n_y=3)}|$ (a,b) and $c_R$ (c,d) for SMAS$_1$ at $g=-1.2$ (a,c) and NRDS$_{II}$ at $g=-3$ (b,d). Solid lines in (a,b) [(c,d)] are reproduced from Fig. 2 (obtained by numerical projection onto the Wannier basis). Open circles are all calculated from the nonlinear Hamiltonian. Insets in (c) and (d) illustrate the underlying mechanisms determining $g_s$ and $g_n$, respectively, with the wiggles depicting the dominant wave dispersion.

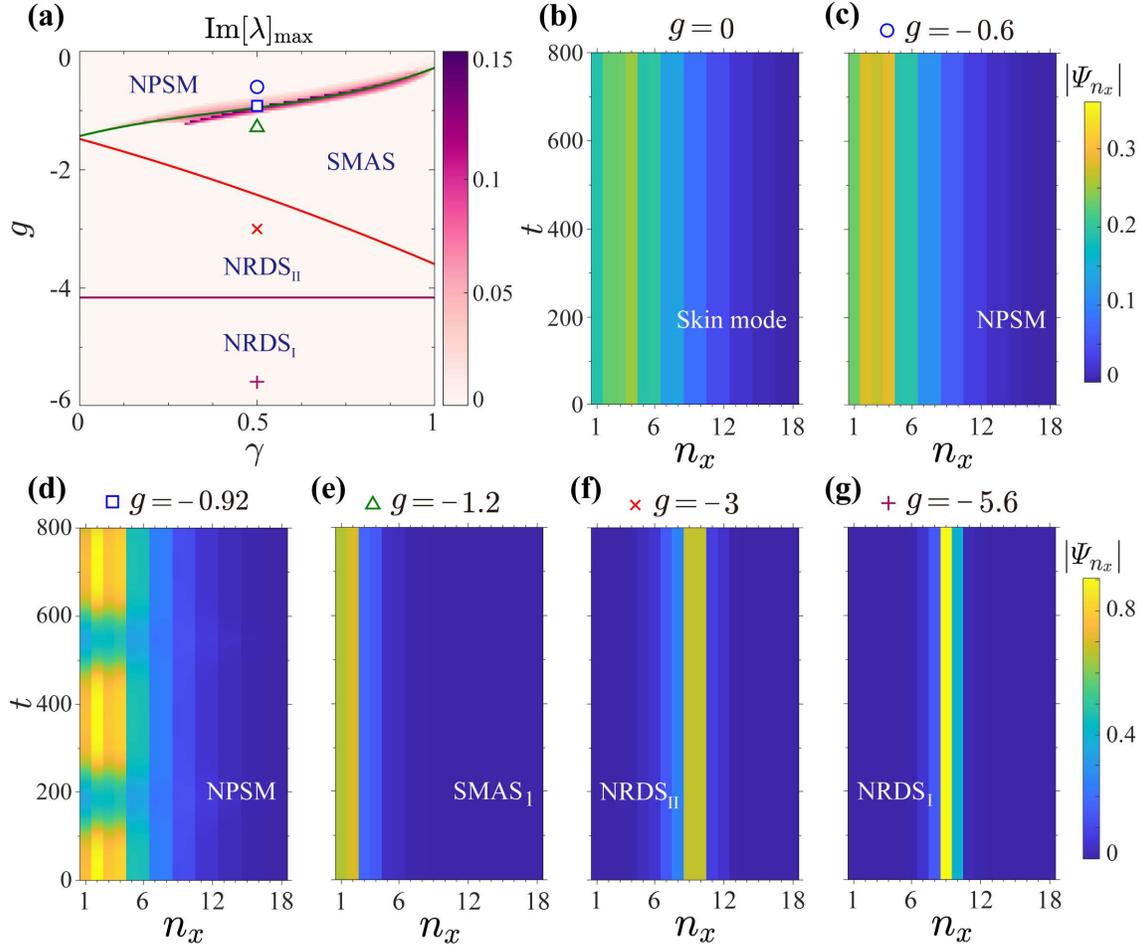

FIG. 5. (a) Stability diagram across the phase diagram of Fig. 3. The color map shows $\text{Im}[\lambda]_{\max}$ for the modes under consideration, while the transition boundaries $g_s$, $g_n$, and $g_t$ are reproduced from Fig. 3. (b)-(g) Time evolution of $|\Psi_{(n_x,n_y=3)}(t)|$ for a stable skin mode (b), a stable NPSM (c), an unstable NPSM (d), a stable $\text{SMAS}_1$ (e), an $\text{NRDS}_{\text{II}}$ (f), and an $\text{NRDS}_{\text{I}}$ (g). (b)-(d) share the upper color bar, while (e)-(g) share the lower one. Their corresponding $g$ values are labelled herein, $\epsilon = 10^{-3}$, and $t$ is in units of $1/2|\kappa|$. The calculations of dynamics are performed under OBCs and using the Runge-Kutta method.

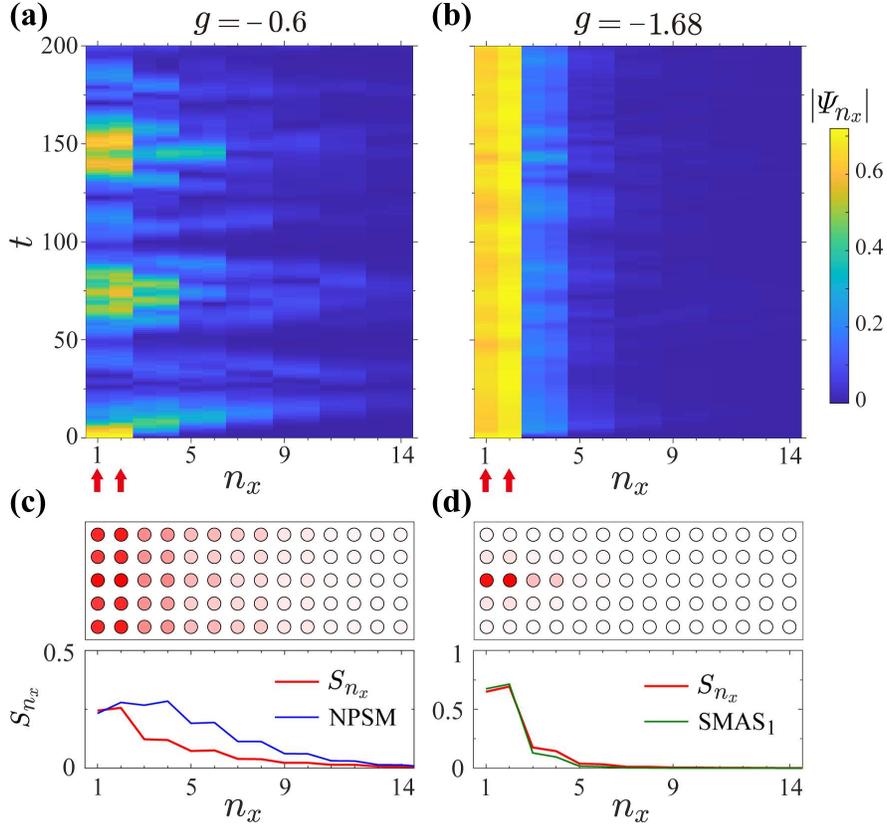

FIG. 6. (a,b) Time evolution of $|\Psi_{(n_x,n_y=3)}(t)|$ for a stable NPSM (a) and a stable SMAS$_1$ (b) under edge excitation. We use two-channel excitation at the favorable boundary, i.e., $\Psi_{\mathbf{n}}(0) = \frac{\sqrt{2}}{2}\delta_{n_x,1}\delta_{n_y,3} + \frac{\sqrt{2}}{2}\delta_{n_x,2}\delta_{n_y,3}$, illustrated by red arrows. $\Psi_{\mathbf{n}}(t)$ is normalized at each time step. (c,d) Spatial distribution of $S_{\mathbf{n}}$ (upper panels) corresponding to (a,b), where color shading represents the magnitude. The lower panels compare $S_{(n_x,n_y=3)}$ (red lines) with $|\psi_{(n_x,n_y=3)}|$ of the nonlinear eigenstates (blue and green lines). We have set $T = 200$ in units of $1/2|\kappa|$. The corresponding $g$ values are labelled herein, and other parameters follow those in Fig. 2.